\documentstyle[preprint,revtex]{aps}
\begin{document}
\preprint{UM-P-93/11}
\preprint{OZ-93/5}
\preprint{Feb. 1993}
\begin{title}
CP Violation in Higgs Decays
\end{title}
\author{Xiao-Gang He, J. P. Ma and Bruce McKellar }
\begin{instit}
Research Center for High Energy Physics\\
School of Physics\\
University of Melbourne \\
Parkville, Vic. 3052 Australia
\end{instit}
\begin{abstract}
We study CP violation in fermion pair decays of Higgs boson. We identify
some CP odd observables related to the tree level
decay amplitude. We find that a few thousand Higgs boson decay events
can already provide important information  about CP violation.
If the Higgs boson is produced, such an analysis could be carried
out at the SSC, LHC and NLC.
\end{abstract}
\newpage
CP violation was found in the neutral
K meson system and can be explained within
the Minimal Standard Model, where the source of CP violation is
the Kobayashi--Maskawa phase\cite{km}.
If this is the only source of CP violation the
observed baryon asymmetry in the universe may not be accommodated \cite{bau},
so additional sources may be needed. The multi-Higgs doublet models,
where CP violation can exist in the Higgs sector\cite{wein},  may provide the
additional sources for CP violation necessary to explain the baryon asymmetry
of the universe\cite{bau}. The possible effects of CP violation from
such models have been studied in
different processes, e.g. the effects in the production
of the $t\bar t$ system were studied in \cite{pes},
the effect in the top quark decay
was studied in \cite{ber,gun}, and
the effects in the neutral Higgs decay were investigated in \cite{dar,son}.
In all the above mentioned works the effects of CP violation come from the one
loop level.

The motivation for our work is to note that
in multi-Higgs doublet models CP violating effects in
fermion pair decays of a neutral Higgs boson exist already at the
{\it tree-level} and hence
can be very large. In this letter we will study these effects in two cases:
a) the Higgs boson is heavy enough to decay into top quark and anti-top
quark, b)  the Higgs is light and the fermion pair is a $\tau$ lepton pair.
Before going into the detailed decay channels, we discuss at first some
general features in  $H\rightarrow f\bar f$. The most general
decay amplitude for this decay is
\begin{equation}
   T_{fi} =\bar f (a_f+i\gamma_5 b_f) f\;,
\end{equation}
where $a_f$ and $b_f$ are in general complex numbers. If both $a_f$ and
$b_f$ are nonzero, CP is violated. To probe  CP violation the polarization
of the fermion pair {\it must} be measured. Since the
fermions we will consider here are top quark and $\tau$ lepton,
the polarization information can be obtained through their decays.
One can define a density matrix $R$
for the process $H\rightarrow f\bar f$, where the $f (\bar f )$ is
polarized and the polarization is described by a unit polarization vector
${\bf n}_{f(\bar f)}$ in the $f(\bar f)$-rest frame. With the amplitude in
eq.(1) the CP violating part of the density matrix is given by
\begin{eqnarray}
   R_{CP}= N_f  \beta_f\{
      {\rm Im}(a_fb_f^*) {\bf\hat p_f}\cdot ({\bf n_{\bar f}}
            -{\bf n_f})
         -{\rm Re}(a_fb_f^*) {\bf\hat p_f}\cdot (
     {\bf n_f}\times {\bf n_{\bar f}})\}\;,
\end{eqnarray}
where $N_f$ is a normalization constant, and
${\bf\hat p_f} $ is the three momentum
direction of the fermion and $\beta_f=\sqrt{1-4m_f^2/M_H^2}$. At the tree-level
$a_f$ and $b_f$ are real and ${\rm Im}(a_fb_f^*)$ is zero. $R_{CP}$
contains all information about experimental observables. The expectation values
of CP odd and CPT odd observables are
proportional to $Im(ab^*)$. To study these
observables one needs to know the absorptive amplitudes $Im a_f$ and $Imb_f$
which
can only be generated at loop levels. Observables of this type has been studied
in \cite{dar}. To
detect possible CP violation at the tree-level one should use CP odd and
CPT even observables which are related to ${\rm Re}(a_fb_f^*)$.
In these observables one naturally expects bigger CP violating
effects
than in CP odd and CPT odd observables. We now study some of these observables
in the two cases mentioned before.
We will neglect the imaginary  amplitudes of $a_f$ and $b_f$.

\noindent
{\it Case a)}. $H\rightarrow t\bar t$. As is well known,
the top quark is heavy and
will then decay quickly before  forming hadronic states. This makes it
possible to analyze the spin of the top quark\cite{bigi}.
We consider the decays
\begin{eqnarray}
 & t \rightarrow W^+ b \rightarrow \ell^+\nu b\;,\nonumber\\
  & \bar t \rightarrow W^- \bar b \rightarrow \ell^-\bar \nu \bar b\;.
\end{eqnarray}

We use the lepton momenta defined in the $W$ rest frames to construct
the CP odd and CPT even observable $O_t$ and the corresponding asymmetry
\begin{eqnarray}
   O_t &=& {\bf \hat p_t}\cdot ({\bf\hat q_+}\times
   {\bf\hat q_-})\;,\nonumber\\
     A_t&=& {N(O_t>0)-N(O_t<0) \over N(O_t>0)+N(O_t<0)}\;,
\end{eqnarray}
where ${\bf\hat q_+}({\bf\hat q_-})$ is the momentum direction of
$\ell^+(\ell^-)$ in the $W^{+(-)}$ rest frame. These momenta are
related to the corresponding momenta measured in the Higgs rest frame
through two Lorenz boosts: one tranforms the top quark into its rest frame,
and the other transforms the $W$ boson into its rest frame. It should be
pointed out that the top quark and the $W$ rest frames
can be reconstructed in experiment even though the neutrinos in eq.(3)
escape and the $b(\bar b)$ quark jet may be not distinguished from
the $\bar b (b)$ quark in experiment\cite{yuan}.
Using the tree level results from the MSM for
the decay matrices of the decays in eq.(3) we obtain
\begin{eqnarray}
   &<&O_t> = -{4\over 9} \beta_t
   {\alpha_t^2 (a_tb_t) \over \beta_t^2  a_t^2
    + b_t^2  }\;,\;\; <O_t^2>={2\over 9}\;,\nonumber\\
   &A_t& = -{ 9\pi\over 16} <O_t>\;,
\end{eqnarray}
where $\alpha_t = M_W(2m_t+M_W)/(m_t^2 + 2M_W^2)$ is the polarization parameter
for $t$.
We also give in eq.(5) the variance $<O_t^2>$ of the observable $O_t$.
The statistical error can then be determined by $\delta <O_t> = \sqrt{
<O_t^2> /N_{event}}$, here $N_{event}$ is the number of the available
events used to measure the observable (note,
$\delta A_t=N_{event}^{-{1\over2}}$). The CP violating effects  can be very
large, several tens of events can give useful information about $x=a_t/b_t$.
For example, if $\vert x\vert$ lies from  0.58 to  3.8 for $M_H=400$GeV
and $m_t=150$GeV, then with one hundred available events the absolute value
of $A_t$ is already larger than $2\cdot \delta A_t$.

We have also worked out one observable which is constructed by the lepton
momenta ${\bf q'_+}$ and ${\bf q'_-}$ measured in the Higgs rest frame
 \begin{eqnarray}
&<&O'_t>=<{\bf \hat p_t}\cdot ( {\bf q'_+}
  \times {\bf q'_-})>
                  = {4\beta_t a_tb_t \over \beta_t^2 a_t^2+b_t^2}
     ( { 1\over 36z}\cdot {z^2+2z+3 \over z+2})^2 \;,\nonumber\\
&<& (O'_t)^2> = 6 M_W^4 ( {z^3+2z^2+3z+4 \over 120z(z+2)})^2\;.
\end{eqnarray}
Here $z=m_t^2/M_W^2$. Comparing the quantities in eq.(4) $<O'_t>$ is less
sensitive to $x$, but, it may be easier to measure than those in eq.(4).

\noindent
{\it Case b)}. $H\rightarrow \tau^-\tau^+$. In this case we use the
following decay mode to analyze the polarization of the $\tau$ leptons
\begin{equation}
 \tau^- \rightarrow \pi^-\nu_{\tau},\ \
      \tau^+ \rightarrow \pi^+\bar \nu_{\tau}\;.
\end{equation}
Denoting ${\bf\hat p_+ }({\bf\hat p_-})$ as the moving direction of the
$\pi^+ (\pi^-)$ in the $\tau^+(\tau^-)$ rest frames, as in case a),
we have
\begin{eqnarray}
&<&O_\tau> = <{\bf\hat p_\tau}\cdot ( {\bf\hat p_+ }
       \times {\bf\hat p_-})> =
      -{4\over 9} \beta_\tau \alpha_\tau^2
   { (a_\tau b_\tau ) \over \beta_\tau^2  a_\tau^2
    + b_\tau^2  }\;,\;\;<O_\tau^2>={2\over 9}\;,\nonumber\\
&A_\tau& ={N(O_\tau >0)-N(O_\tau<0) \over
         N(O_\tau >0)+N(O_\tau<0)}=
     -{ 9\pi\over 16} <O_\tau>\;.
\end{eqnarray}
Here $\alpha_\tau = 1$.
One can also use the leptonic decay mode of the $\tau$ leptons instead of
using the decays in eq.(8). In this case
$\alpha_\tau = 1/3$.

The momenta of the $\tau$ leptons are difficult to measure, so it is also
difficult to reconstruct the $\tau$ rest frame. It may be possible to
overcome this difficulty by constructing CP odd correlations between
the momenta measured in the laboratory  system and any beam direction
if the Higgs boson is produced at some $e^+e^-$ or $p\bar p$ colliders.

The above results can also be applied to $H\rightarrow \mu^-\mu^+$.
In this case
the polarization of muon is analysed by its leptonic decay with $\alpha_\mu
=1/3$.

It is clear from the formula for $A_f$ that the asymmetry can be of
order one if $a_f$ and $b_f$ are about the same strength. There is experimental
constraint for the parameter $a_fb_f$ from the neutron electric dipole
moment measurement. However, at the present the constraint is not very strong
\cite{xgh}.
The maximal value for $A_{max} =\alpha_i^2\pi/8$ is not ruled out.
The parameters $a_f$ and $b_f$ can receive nonzero contributions from some
models at the tree level.
Let us consider two Higgs-doublet models.
In these models the gauge group is the same as the MSM, i.e.,
$SU(3)_C\times SU(2)_L\times U(1)_Y$ and
has two Higgs representations transforming under $SU(2)_L$ as doublet.
It is possible to have CP violation in the Higgs
sector in these models, in  which there
are three physical neutral ($H_j$ mass eigenstates) and one charged physical
Higgs bosons.
In order to prevent large flavour changing neutral currents at the tree level,
some discrete symmetries are imposed to the Yukawa sector and the choice
of the discrete symmetry is not unique. Different discrete symmetries
result in different Yukawa interaction. A possible Yukawa nteraction
Lagrangian is
\begin{eqnarray}
L &=& (\sqrt{2}G_F)^{1/2}[\bar U_i U_i m_{U_i}(\mbox{d}_{1j}-\mbox{cot}\beta
\mbox{d}_{2j})
+i\bar U_i \gamma_5 U_i m_{U_i} \mbox{cot}\beta \mbox{d}_{3j}\nonumber\\
& & + \bar D_i D_i m_{D_i}(\mbox{d}_{1j} + \mbox{tan}\beta \mbox{d}_{2j})
+ i\bar D_i \gamma_5
D_i m_{D_i} \mbox{tan}\beta \mbox{d}_{3j}\\
& & + \bar L_i L_i (\mbox{d}_{1j} + \mbox{tan}\beta \mbox{d}_{2j}) +
i\bar L_i\gamma_5 L_i
m_{L_i} \mbox{tan}\beta \mbox{d}_{3j}] H_j\nonumber\;,
\end{eqnarray}
where $U_i$, $D_i$ and $L_i$ are the up-, down-quarks and charged leptons
respectively, the sub-indices $i$ runs for different generations and
$j$ runs for the
three different neutral Higgs particles, $\mbox{d}_{ij}$ are the
mixing angles of
the Higgs mass matrix, and $\mbox{tan}\beta$ is the ratio of the
vacuum expectation
values of the two Higgs doublets. If $\mbox{d}_{1j}\mbox{d}_{3j}$ and
$\mbox{d}_{2j}\mbox{d}_{3j}$ are
non-vanishing, CP is violated. We can easily read off the parameters
$a_f$ and $b_f$ from the above Lagrangian.

Without loss of generality, let us assume that $H_1$ is the Higgs boson to be
produced and its decays to be analyzed.
For $H\rightarrow t\bar t$ we have
\begin{eqnarray}
a_t &=& -(\sqrt{2}G_F)^{1/2}m_t(\mbox{d}_{11} - \mbox{cot}\beta \mbox{d}_{21})
\;,\nonumber\\
b_t &=& -(\sqrt{2}G_F)^{1/2}m_t \mbox{cot}\beta \mbox{d}_{31}\;.
\end{eqnarray}
In the region where $M_H > 2m_t$, if one assumes that the MSM prediction
is roughly correct for the branching ratios for the Higgs decay,
then the main
decay modes are $H_1\rightarrow W^+W^-,\ ZZ$. Nevertheless,
$H_1\rightarrow t\bar t$ also
has a substantial branching ratio, i.e., for $m_t =150GeV$ and $M_H = 400 GeV$,
$B_t(H_1\rightarrow t\bar t)$ is about 0.17. Taking the semi-leptonic
branching ratios of the top quark decays into account, CP violation
may be observed at $90\%$ confidence level (here we only considered statistic
error) with 6,000
Higgs bosons. However, in the model we consider, $\mbox{d}_{11}$
may be very small,
and the decay rate for $H_1\rightarrow W^+W^-,\ ZZ$ is proportinal
to $\mbox{d}_{11}^2$. In this case, the branching ratio for
$H_1\rightarrow t\bar t$
may be large. It is possible to observe CP violation with less than one
thousand Higgs bosons.

In the range where $M_H < 2M_W$,
the main decay modes is $H\rightarrow b\bar b$.
Because the b quark forms hadrons before it can decay through weak
interactions the information about the polarization of the b quark is washed
out, thus only the leptonic decay channels can be used for CP test.
{}From eq.(9) we find for the $\tau$ leptons
\begin{eqnarray}
a_\tau &=& -(\sqrt{2}G_F)^{1/2}m_\tau(\mbox{d}_{11}+\mbox{tan}\beta \mbox{d}
_{21})\;,\nonumber\\
b_\tau &=& -(\sqrt{2}G_F)^{1/2}m_\tau \mbox{tan}\beta \mbox{d}_{31}\;.
\end{eqnarray}
In this
case the branching ratio for Higgs to $\tau$ pair is
\begin{equation}
B_\tau(H_1\rightarrow \tau\bar\tau) \approx {m_\tau^2\over m_\tau^2 + 3m_b^2}
\approx 0.04\;.
\end{equation}
and $B(\tau \rightarrow \pi^-\nu_\tau)\approx 11\%$.
About $6\times 10^{4}$ Higgs bosons are required
to see the maximal CP violation.
However, as pointed out earlier, the Yukawa interaction can be different
from those
we are discussing, the branching ratio may be large.

In the range $M_H >2M_Z$, if the MSM prediction for Higgs decay branching
ratio is roughly correct, then the branching ratio for $H_1 \rightarrow
\tau^+\tau^-$ is very small. However, the situation here is similar as
discussed for $H_1 \rightarrow t \bar t$, it is possible in two Higgs doublet
models to have larger branching ratio for $H_1 \rightarrow \tau^+\tau^-$. This
process may still be a good place to test CP invariance in this region.

The above analysis can be carried out for $H_1 \rightarrow \mu \bar \mu$. Of
course the branching ratio for this decay is much smaller, $B(H_1 \rightarrow
\mu \bar \mu)\approx 10^{-4}$. We need about $10^{6}$ Higgs bosons to see
CP violation. This is still achievable at SSC.

The above discussion can be easily genaralized to multi-Higgs models.

In conclusion, we have studied CP violation in $H\rightarrow f\bar f$. We
have proposed the study of CP odd observables to which CP violation at
the tree level
can contribute. The predictions for the observables are worked out,
and two Higgs doublet models are discussed.
Less than one thousand Higgs bosons
may  provide important information about CP violation. The analysis discussed
in the above can be easily carried out at the SSC, LHC, and NLC.

\acknowledgments
This work was supported in part by the Australian Research Council. XGH and
JPM thank Dr. S. Torvey for useful discussions.

\end{document}